\documentstyle[a4,aps,graphicx]{revtex}

\tighten
\draft

\preprint{\vbox{TPJU-3/2002}}

\begin{document}
\title{~~~~~~~~~~~~~~~~~~~~~~~~~~~~~~~~~~~~~~~~~~~~~
~~~~~~~~~~~~~~~~~~~~~~TPJU 3/2002\\
~~\\
Pion and vacuum properties
in the nonlocal NJL model}

\author{Micha{\l} Prasza{\l}owicz\footnote{Presented
at the 8-th Adriatic Meeting, Sept. 4-14, 2001, Dubrovnik,
Croatia} and Andrzej Rostworowski}
\address{~\\
Institute of Physics, Jagellonian University, \\
ul. Reymonta 4, 30-059 Krak{\'o}w, Poland}

\maketitle

\begin{abstract}
 We formulate the nonlocal NJL model with a
momentum dependent constituent quark mass and calculate pion light
cone wave functions of twist 2 and 3. The leading twist wave
function is not asymptotic and agrees well with the new CLEO data.
Normalization conditions for the twist 3 wave functions are used
to calculate the quark condensate. A prescription to calculate the
gluon condensate is proposed. The numerical value of the gluon
condensate nicely agrees with the phenomenological value, whereas
the quark condensate is larger than the phenomenological value of
$-(250$ MeV$)^3$. The relation between the  $k_T^2$ moments and
mixed condensates are used to estimate the mixed quark-gluon
condensate of dimension 5.
\end{abstract}

\section{Introduction}

\label{s:intro}

In this short note we shall describe a simple and tractable model
for the pion light cone wave functions which is based on the
instanton model of the QCD\ vacuum. Hadron light cone wave
functions were theoretically introduced more than 20 years ago
\cite{ChZh}\nocite{BrLep,FarJack,EfRad}-\cite{ChZhitZhit}.
Recently the analysis of Ref.\cite{YakSch} based on the latest
CLEO measurements \cite{Cleo} put some limits on the expansion
coefficients of the axial-vector (AV) pion wave function in terms
of the Gegenbauer polynomials. This analysis indicates that the
pion wave function measured at
$Q^{2}=1.5-9.2$ GeV$^{2}$ is neither asymptotic $\phi _{\mathrm{as}%
}^{AV}(u)=6\;u(1-u)$ (with $u$ being the fraction of the pion momentum carried
by the quark) nor of the form proposed by Chernyak and Zhitnitsky in 1977
\cite{ChZhPRep}: $\phi _{\mathrm{CZ}}^{AV}(u)=30\;u(1-u)(1-2u)^{2}$. These two
wave functions together with a typical prediction of the present model are
shown in Fig.1a. In Fig. 1b we show the 95\% and 68\% confidence level
contour plots in the $a_{2}-a_{4}$ parameter space from the analysis of
Schmedding and Yakovlev (Fig.6 in Ref.\cite{YakSch}) together with the
values of $a_{2}$ and $a_{4}$ for $\phi _{\mathrm{as}}^{AV}$, $\phi _{%
\mathrm{CZ}}^{AV}$ and various parameters of the present model.

The instantanton model, after
integrating out gluons and performing the bosonization, reduces to
a simple Nambu-Jona Lasinio type model where the quarks interact \emph{%
nonlocally} with an external meson field $U$ \cite{DPrev,DP}:
\begin{equation}
S_{I}=\int \frac{d^{4}kd^{4}l}{(2\pi )^{8}}\bar{\psi}(k)\sqrt{M(k)}U^{\gamma
_{5}}(k-l)\sqrt{M(l)}\psi (l)\,  \label{SI}
\end{equation}
and $U^{\gamma _{5}}$ can be expanded in terms of the pion fields:
\begin{equation}
U^{\gamma _{5}}=1+\frac{i}{F_{\pi }}\gamma ^{5}\tau ^{A}\pi ^{A}-\frac{1}{%
2F_{\pi }^{2}}\pi ^{A}\pi ^{A}+\ldots  \label{U}
\end{equation}
Here $F_{\pi }=93$ MeV and $M(k)=MF^{2}(k)$ is a momentum
dependent constituent quark mass which also plays a role of the
pion-quark coupling. Let us note that in the instanton model both quark
and gluon condensation occur at the same scale $\mu_0 $ which
is associated with the average instanton size $1/\overline{\rho }=600$~MeV.

In principle $F(k)$ has been calculated in
the instanton model in the Euclidean space time. Here, following
Refs.\cite{Bochum,Praszalowicz:2001wy}, we will perform
calculations directly in the Minkowski space. To this end we shall
choose a simple pole formula \cite{Praszalowicz:2001wy}
\begin{equation}
F(k)=\left( -\frac{\Lambda ^{2}}{k^{2}-\Lambda ^{2}+i\epsilon }\right) ^{n}
\label{cutform}
\end{equation}
which for $n\sim 2-3$ and for $k^{2}<0$ reproduces the $k$ dependence
obtained from the instantons reasonably well \cite{Praszalowicz:2001wy}
(see Fig. 2a).
Here $M=M(0)$ is a model parameter which we choose to be of the order of
$350 $ MeV.

\begin{figure}[h]
\centerline{\includegraphics[width=.5\textwidth]{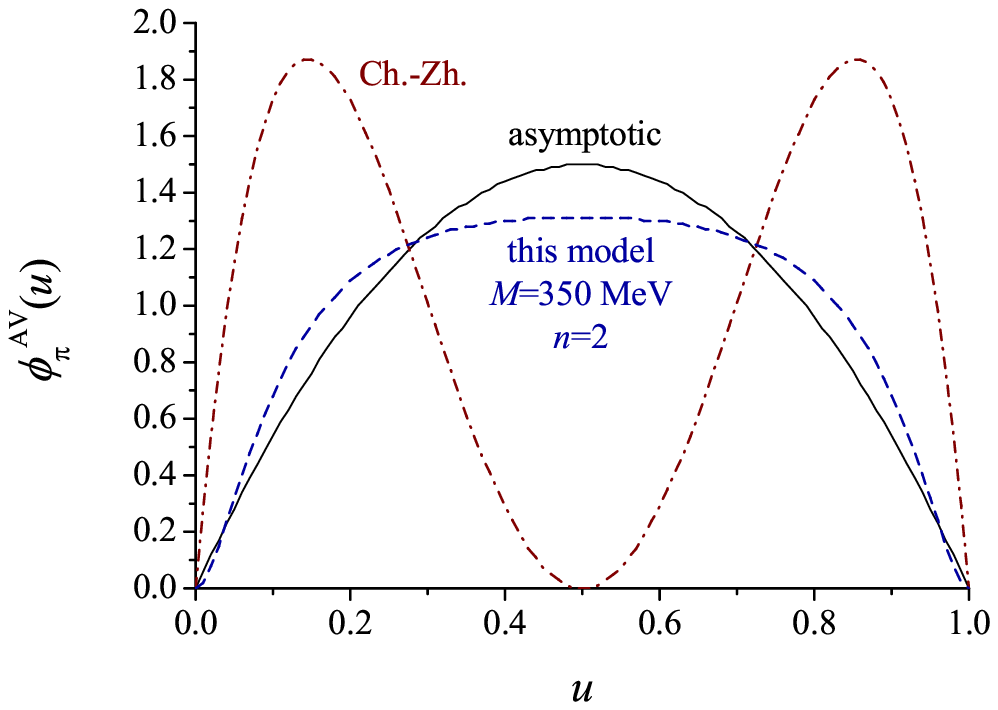}
\includegraphics[width=.5\textwidth]{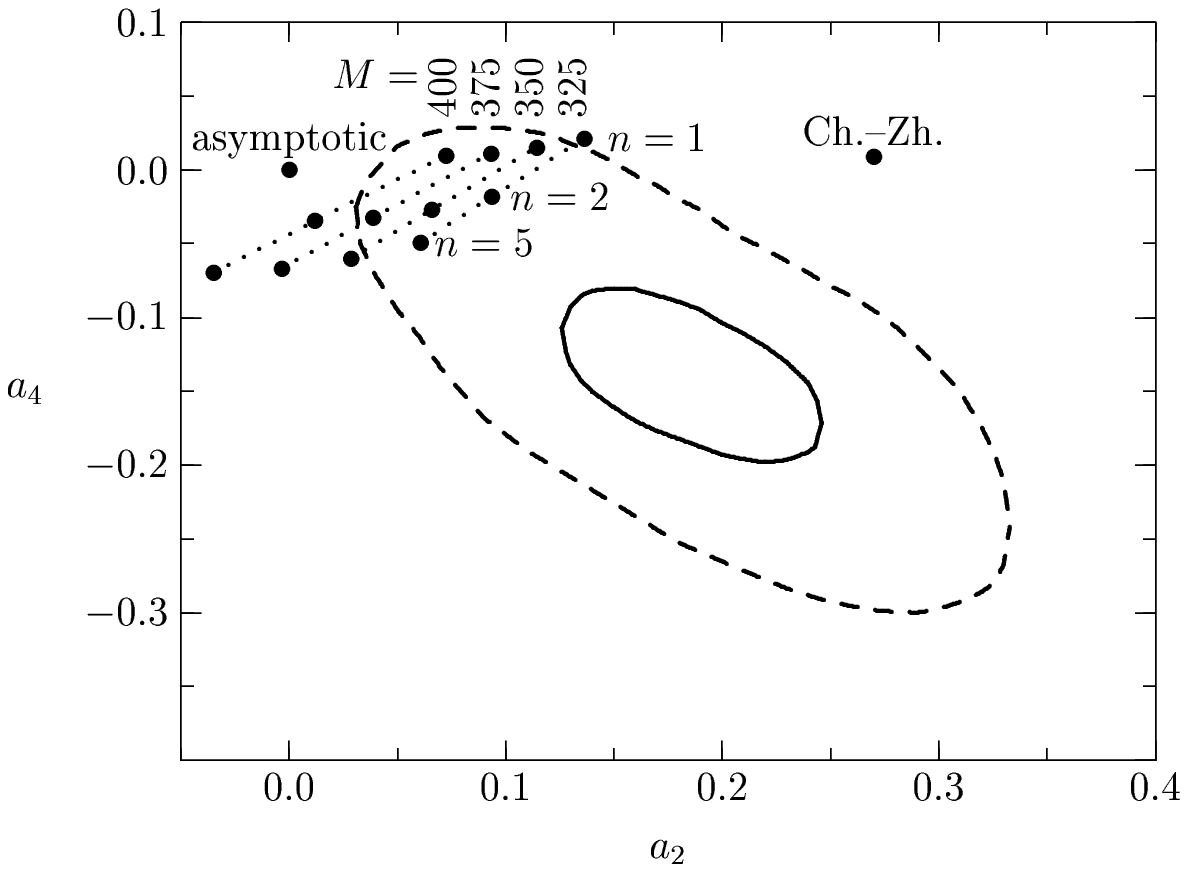}}
\caption[]{Left: Asymptotic and Chernyak-Zhytnitsky
leading twist pion wave functions together with a typical wave
function from the present model. 
Right: The parameter space
($a_{2},a_{4}$) of Ref.[6]. Black dots represent different model
predictions, solid contour corresponds to 68\% confidence level,
whereas the dashed one to 95\%}
\label{eps1}
\end{figure}

As we shall see, the model is technically very simple and allows
to calculate pion wave functions (not only the axial-vector, but
also the pseudo-scalar (PS) and the pseudo-tensor (PT) ones)
analytically up to a numerical solution of a certain algebraical
equation of the order $4n+1$. Given this simplicity it is of
importance to perform various tests in order to gain confidence in
the model as well as to find its limitations. In this paper we
provide 4 kinds of tests.

First we calculate the leading twist pion wave function and
compare with the existing data. Next we calculate the non-leading
twist wave functions, which are normalized to the quark
condensate. This allows us to calculate $\left\langle
\bar{q}q\right\rangle $.

It is important to note that in our approach we calculate not only the $u$
dependence but also the dependence on the transverse momentum $k_{T}$:
\begin{equation}
\phi _{\pi }(u)=\int\limits_{0}^{\infty }dk_{T}^{2}\,\psi _{\pi
}(u,k_{T}^{2}), \;\;\;\;\;
\tilde{\phi}_{\pi }(k_{T}^{2})=\int\limits_{0}^{1}du\,\psi _{\pi
}(u,k_{T}^{2}).  \label{dkt}
\end{equation}
By calculating $k_T^2$ moments we get the mixed condensate of
dimension 5.

Another advantage of our method is that the analytical expression
for the quark condensate is given in terms of a Minkowskian
integral which in a limit of a constant $M(k)$ and
$k^{2}\rightarrow -k_{E}^{2}$ reduces to the well known Euclidean
form. By comparing the two expressions one can by inspection guess
a continuation prescription which allows to rewrite certain
Euclidean integrals as the Minkowskian ones. We use this in some
respect \emph{ad hoc} prescription to calculate the gluon
condensate $\left\langle \alpha /\pi \,GG\right\rangle $, which
provides another test of our approach.

\section{Pion wave functions in the nonlocal quark model}

\label{s:cond}

We shall be dealing with the leading twist axial-vector (AV),
twist 3 pseudo-scalar (PS) and pseudo-tensor (PT) wave functions
defined as follows \cite{BrauFil2,Ball:1999je}:
\begin{eqnarray}
\phi _{\pi }^{AV}(u)& =&\frac{1}{i\sqrt{2}F_{\pi }}\int\limits_{-\infty
}^{\infty }\frac{d\tau }{\pi }e^{-i\tau (2u-1)(nP)}\left\langle 0\right|
\bar{\psi}(n\tau )\rlap{/}n\gamma _{5}\psi \left( -n\tau \right) \left| \pi
^{+}(P)\right\rangle , \label{Phidefs} \\
\phi _{\pi }^{PS}(u)& =&-(nP)\frac{F_{\pi }}{\sqrt{2}\left\langle \bar{q}%
q\right\rangle }\int\limits_{-\infty }^{\infty }\frac{d\tau }{\pi }e^{-i\tau
(2u-1)(nP)}\left\langle 0\right| \bar{\psi}(n\tau )i\gamma _{5}\psi \left(
-n\tau \right) \left| \pi ^{+}(P)\right\rangle ,  \nonumber \\
\phi _{\pi }^{PT}(u)& =&
\frac{-6F_{\pi }}{\sqrt{2} \left\langle \bar{%
q}q\right\rangle }\int\limits_{0}^{u}dw\int\limits_{-\infty
}^{\infty }\frac{d\tau}{\pi} e^{-i\tau (2w-1)(nP)}
n^{\alpha }P^{\beta }\left\langle 0\right| \bar{\psi}%
(n\tau )\sigma _{\alpha \beta }\gamma _{5}\psi \left( -n\tau
\right) \left| \pi ^{+}(P)\right\rangle.  \nonumber
\end{eqnarray}
where we have chosen $n=(1,0,0,-1)$ as a light-cone vector parallel to $%
z_{\mu }=\tau n_{\mu }$ and $\tilde{n}=(1,0,0,1)$ parallel to $P_{\mu}$.
All three
wave functions are normalized to 1. The normalization condition for $\phi
_{\pi }^{PS}$ and yield $\phi _{\pi }^{PS}$ therefore the expression for $%
\left\langle \bar{q}q\right\rangle $, whereas normalization of $\phi _{\pi
}^{AV}$ is used to fix the model parameter $\Lambda $ for given $M$ and $n$.

Technically speaking all three wave functions (\ref{Phidefs}) are
given in terms of a loop integral with a momentum dependent quark
mass $M(k)$, which also acts as a quark-pion coupling. In order to
calculate the loop integral we have to find zeros of the
propagators which are generically of the form
\begin{equation}
k^{2}-M^{2}\left[ \frac{\Lambda ^{2}}{k^{2}-\Lambda ^{2}+i\epsilon }\right]
^{4n}+i\epsilon =0.  \label{prpopzero}
\end{equation}
Equation (\ref{prpopzero}) can be conveniently rewritten as:
\begin{equation}
z^{4n+1}+z^{4n}-\mu ^{2}=0  \label{zeq}
\end{equation}
where $z=k^{2}/\Lambda ^{2}-1+i\epsilon $ and $\mu
^{2}=M^{2}/\Lambda ^{2}$. In the light cone parametrization
$d^{4}k=1/2\,dk^{+}dk^{-}d^{2}\vec{k}_{T}$ where $k^{\mu
}=k^{+}\tilde{n}^{\mu }/2+k^{-}n^{\mu }/2+k_{T}^{\mu }$. Since
$k^{+}=uP^{+}$ is fixed, equation (\ref{prpopzero}) should be
understood as an equation for $k^{-}$. Generally, equation
(\ref{zeq}) has $4n+1$ complex solutions which in the following
will be denoted as $z_{i}$. These solutions depend on the specific
value of $\mu ^{2}$ and have to be calculated numerically.

Here one faces immediately the problem how to choose the integration contour
in the complex $k^{-}$ plane. The prescription is very simple and has been at
length discussed in Ref.\cite{Praszalowicz:2001wy}. As a result the $dk^{-}$
integrals yield real wave functions which vanish for $u$ outside the region $%
0<u<1$. Moreover for $\Lambda \rightarrow \infty $, i.e. for a constant $%
M(k) $, this prescription reduces in a continuous way to the standard one of
Feynman.

With this prescription the calculations are rather straightforward and we
obtain:
\begin{equation}
\psi _{\pi }^{AV}(u,k_{T}^{2})=\frac{N_{c}}{(2\pi )^{2}}\frac{M^{2}}{\Lambda
^{2}F_{\pi }^{2}}\sum\limits_{i,k=1}^{4n+1}f_{i}f_{k}\,\frac{%
uz_{i}^{n}z_{k}^{3n}+(1-u)z_{i}^{3n}z_{k}^{n}}{t+1+uz_{i}+(1-u)z_{k}},
\label{AVukt}
\end{equation}
\begin{equation}
\psi _{\pi }^{PS}(u\,,k_{T}^{2})=\frac{N_{c}}{(2\pi )^{2}}\frac{M}{%
\left\langle \bar{q}q\right\rangle }\sum\limits_{i,k=1}^{4n+1}f_{i}f_{k}%
\frac{z_{i}^{3n}z_{k}^{3n}(1+\frac{z_{i}+z_{k}}{2})-\mu
^{2}z_{i}^{n}z_{k}^{n}}{t+1+uz_{i}+(1-u)z_{k}},  \label{PSukt}
\end{equation}
\begin{equation}
\psi _{\pi }^{PT}(u\,,k_{T}^{2})=\frac{3N_{c}}{\left( 2\pi \right) ^{2}}%
\frac{M\Lambda ^{2}}{\langle \overline{q}q\rangle }\sum%
\limits_{i,k=1}^{4n+1}f_{i}f_{k}z_{i}^{3n}z_{k}^{3n}\ln
(1+t+uz_{i}+(1-u)z_{k}).  \label{PTukt}
\end{equation}
(where $t=k_{T}^{2}/\Lambda ^{2}$). Factors $f_{i}$ obey the following
properties:
\begin{equation}
f_{i}=\prod\limits_{_{\scriptstyle k\neq i}^{\scriptstyle k=1}}^{4n+1}\frac{1%
}{z_{i}-z_{k}},\;\;\;
\sum\limits_{i=1}^{4n+1}z_{i}^{m}f_{i}=\left\{
\begin{array}{ccc}
0 & {\rm for} & m<4n \\
&  &  \\
1 & {\rm for} & m=4n
\end{array}
\right.  \label{fprop}
\end{equation}
which are crucial for the convergence of the $dt$ integrals.

It is now straightforward to perform either the $dt$ integration
in order to get $\phi _{\pi }$, or the $du$ integration to get the
$k_{T}$ -dependent functions $\tilde{\phi}_{\pi }$.

\section{Properties of the pion wave functions}

\begin{figure}[t]
\centerline{\includegraphics[width=.4\textwidth]{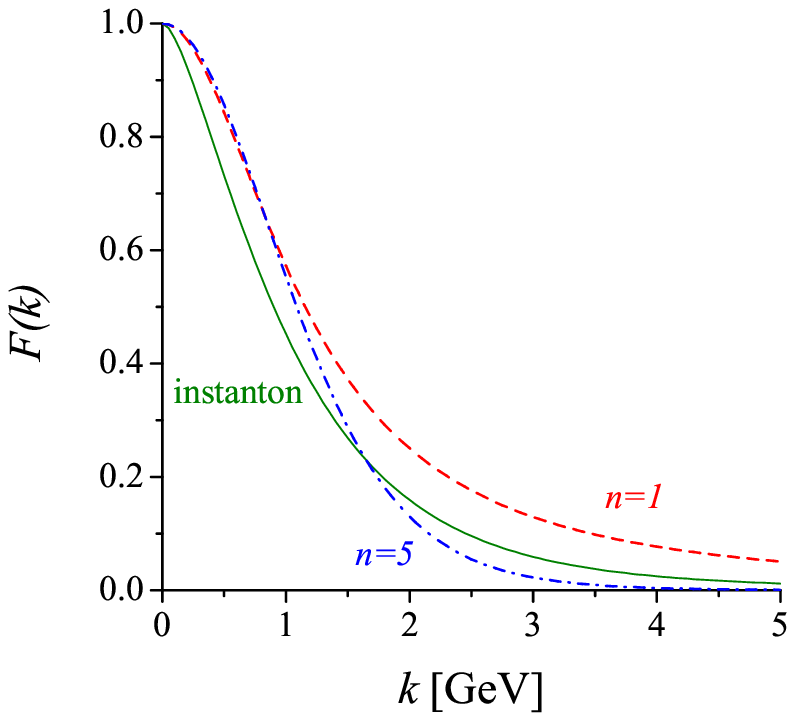}
\includegraphics[width=.5\textwidth]{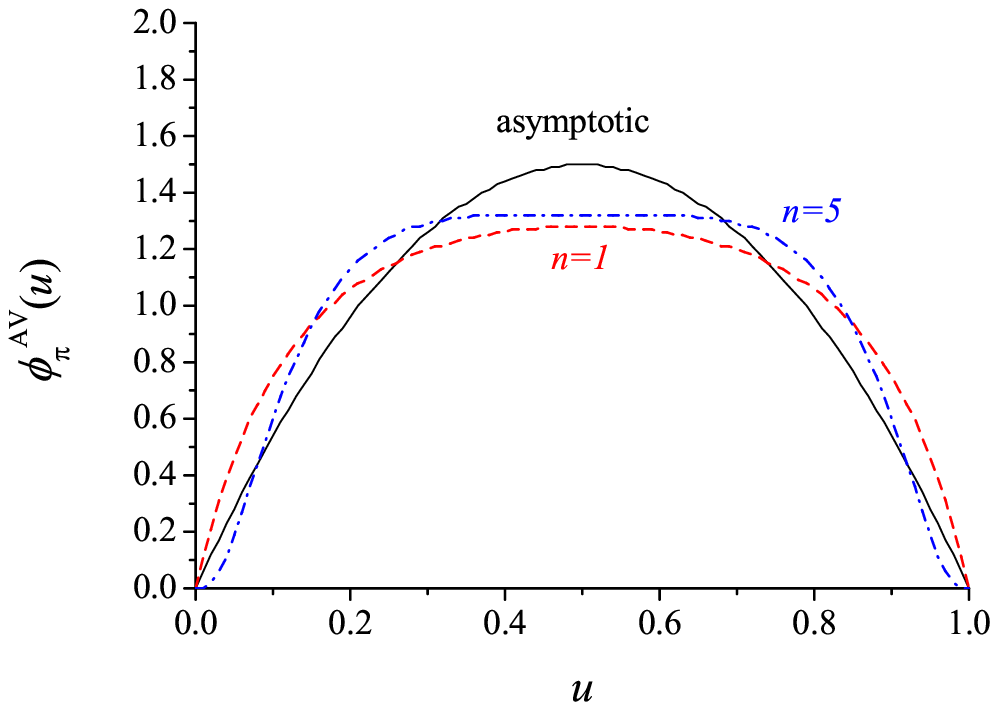}}
\caption[]{Left: $F(k)$ for
Euclidean momentum $k^{2}<0$, for $n=1$ (dashed), 5 (dashed-dotted)
and for the instanton model (solid). 
Right: Axial-vector pion wave function for $M=350$~MeV and for
$n=1$ (dashed) and 5 (dashed-dotted) together with the
asymptotic one (solid)}
\label{eps2}
\end{figure}

In order to study the model dependence on the choice of $M$ and $n$ we have
calculated pion wave functions for $M=325-400$ MeV and $n=1-5$.  The cutoff
parameter $\Lambda $ was adjusted by imposing the normalization condition on
$\phi _{\pi }^{AV}$. In fact, as discussed in Ref.\cite{Praszalowicz:2001wy}%
, the leading twist pion wave function $\phi _{\pi }^{AV}(u)$ does
not change any more if we increase $n$ above $5$. On the other
hand for $n > 5$ the cutoff function (\ref{cutform}), if continued
to the Euclidean metric, starts to deviate significantly from the
one obtained in the instanton model. Therefore we have chosen to
work with $n_{{\rm max}}=5$. In Figs. 2b and 3 we have plotted
$\phi _{\pi }^{AV}$, $\phi _{\pi }^{PS}$ and $\phi _{\pi }^{PT}$
for $M=350$ MeV and $n=1,5$.

\begin{figure}[b]
\centerline{\includegraphics[width=.5\textwidth]{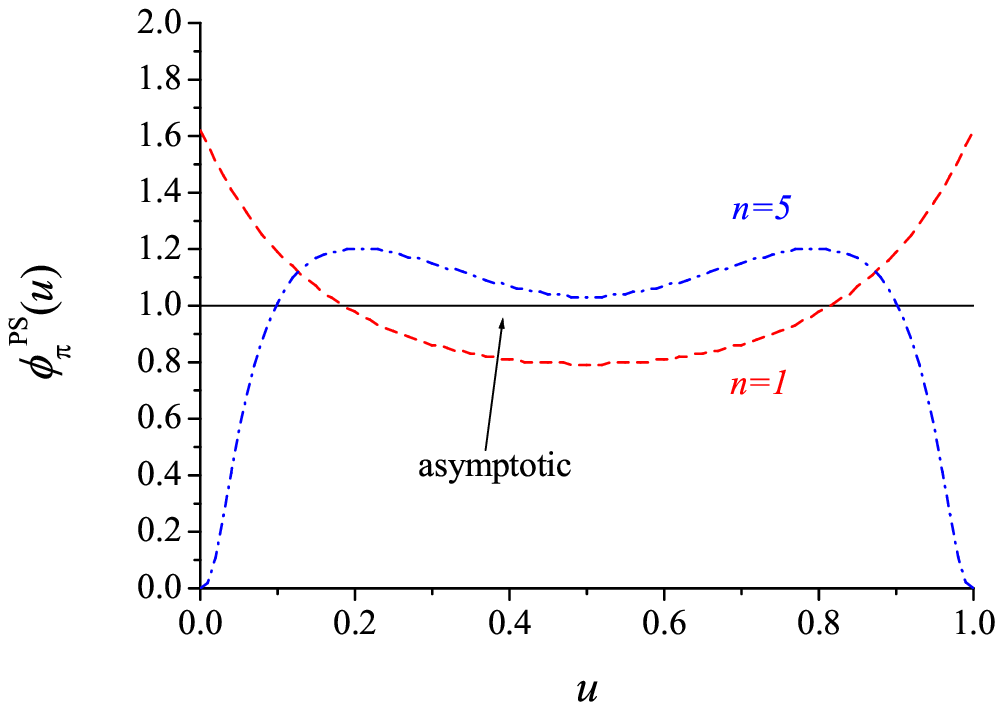}
\includegraphics[width=.5\textwidth]{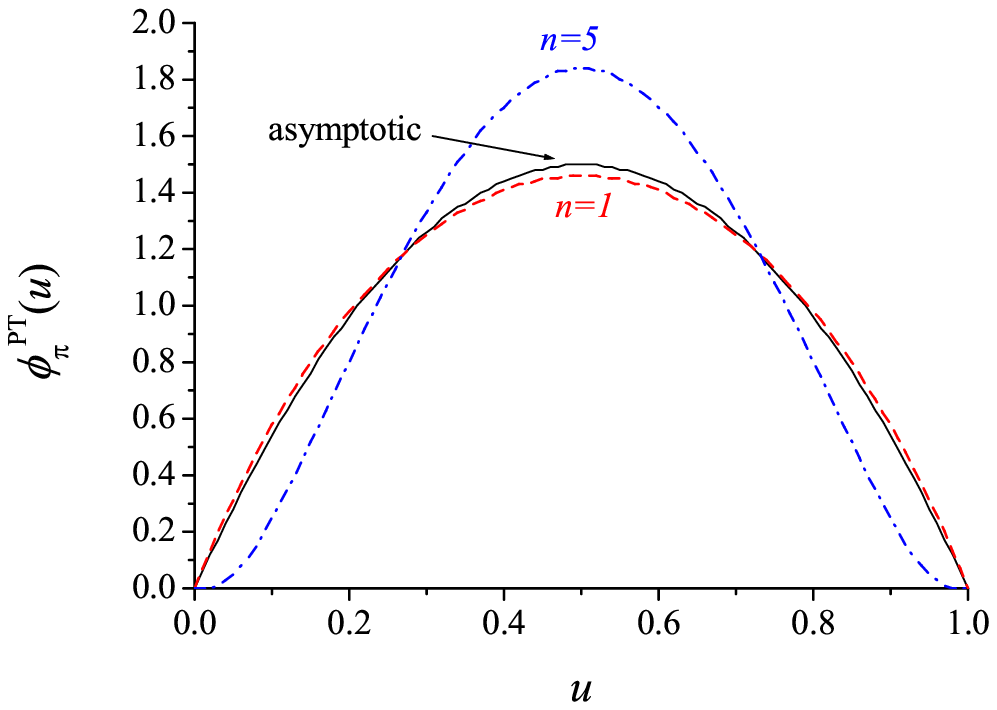}}
\caption[]{Pseudo scalar (left panel) and pseudo-tensor (right panel)
pion wave function for $M=350$~MeV and for $n=1$ (dashed) and 5 (dashed-dotted)
together with the asymptotic one (solid)}
\label{eps3}
\end{figure}

Let us shortly summarize our findings. The axial-vector wave function, $\phi
_{\pi }^{AV}$, vanishes at the end points as $u^{n}$ (or $(1-u)^{n}$) and
shows a plateau around $u=0.5$ with a small dip for $n=5$. It differs from
the asymptotic wave function $\phi _{\mathrm{as}}^{AV}$ and, as seen from
Fig. 1b, the best agreement with the recent analysis of the CLEO data is
obtained for $M=325$ MeV and $n=2-5$ or $M=350$ MeV and $n=2$. The fact
that the true pion distribution amplitude may
be broader than the asymptotic one has been already pointed out in Ref.\cite
{NiSKim}. Such a behavior was found then in Ref.\cite{BaMiSte} where not only
the nonlocality (within the sum rules approach) but also the radiative
corrections have been taken into account.

The pseudo-scalar pion wave function, $\phi _{\pi }^{PS}$, was calculated
within the QCD sum rules in Refs.\cite{BrauFil2,Ball:1999je}. It had a $u-$
shape and did not vanish at the end points. In our case $\phi _{\pi }^{PS}$
vanishes at the end points for $n>1$. Indeed
\begin{equation}
\phi _{\pi }^{PS}(1)\sim \sum\limits_{i}f_{i}\ln
(1+uz_{i})\sum\limits_{k}f_{k}\left[ z_{i}^{3n}z_{k}^{3n}+\frac{1}{2}%
z_{i}^{3n+1}z_{k}^{3n}+\frac{1}{2}z_{i}^{3n}z_{k}^{3n+1}-\mu
^{2}z_{i}^{n}z_{k}^{n}\right]   \label{endp}
\end{equation}
(and similarly for $u=0$) which is equal to 0 due to the property (\ref
{fprop}) except for $n=1$ where $3n+1=4n$. Interestingly, the vanishing of $%
\phi _{\pi }^{PS}$ for $u=0,1$ is correlated with the nonconvexity
of $\phi _{\pi }^{AV}$ at the end points, which, as stated above,
behaves like $u^{n}$ (or $(1-u)^{n}$) for $u\rightarrow 0$ (or
$1$). In any case $\phi _{\pi }^{PS}$differs from its asymptotic
form $\phi _{\mathrm{as}}^{PS}\equiv 1$.

Both pseudo-scalar and pseudo-tensor wave functions show stronger $n$
dependence than $\phi _{\pi }^{AV}$. For $n=1$ $\phi _{\pi }^{PT}$ coincides
with the asymptotic expression $\phi _{\mathrm{as}}^{PT}=\phi _{\mathrm{as}%
}^{AV}$, while for $n=5$ its is depleted at the end points and peaked in the
center.

\section{Condensates}

Since the model parameters are fixed by the normalization of the
axial-vector wave function we could use the normalization condition for $%
\phi _{\pi }^{PS}$or $\phi _{\pi }^{PT}$ to calculate the quark condensate.
The results are presented in Table 1. We see that the quark condensates
obtained from the two normalization conditions do not coincide. In fact,
for almost all model parameters considered, we find
\begin{equation}
\sqrt[3]{\left\langle \bar{q}q\right\rangle _{PS}/\left\langle \bar{q}%
q\right\rangle _{PT}}\simeq 0.9.
\end{equation}
In absolute values the quark condensate calculated within our model
overshoots the phenomenological value of $-(250$ MeV$)^{3}$. This is mostly
due to the rather poor convergence of the $dk_{T}^{2}$ integrals (\ref{dkt}%
). Indeed for large $k_{T}^{2}$:
\begin{equation}
\tilde{\phi}_{\pi }^{AV}(k_{T}^{2})\sim \left( \frac{1}{k_{T}^{2}}\right)
^{4n+1},\;\tilde{\phi}_{\pi }^{PS}(k_{T}^{2})\sim \left( \frac{1}{k_{T}^{2}}%
\right) ^{2n},\;\tilde{\phi}_{\pi }^{PT}(k_{T}^{2})\sim \left( \frac{1}{%
k_{T}^{2}}\right) ^{2n}.
\end{equation}
This is also the reason of rather strong $n$ dependence of $\phi _{\pi
}^{PS}$and $\phi _{\pi }^{PT}$.

The Euclidean formula for the gluon condensate in the instanton model of the
QCD vacuum reads \cite{DPrev}:
\begin{equation}
\left\langle \frac{\alpha }{\pi }GG\right\rangle =32N_{c}\int \frac{%
d^{4}k_{E}}{(2\pi )^{4}}\frac{M^{2}(k_{E})}{k_{E}^{2}+M^{2}(k_{E})}.
\label{GGcond}
\end{equation}
Apart from the numerical factor in front it differs from $\left\langle \bar{%
q}q\right\rangle $ by an additional power of $M(k_{E})$ in the numerator. In
Ref.\cite{Praszalowicz:2001wy} we have suggested the continuation prescription
of (\ref{GGcond}) to the Minkowski metric with the result
\begin{equation}
\left\langle \frac{\alpha }{\pi }GG\right\rangle =-\frac{8N_{c}M^{2}\Lambda
^{2}}{(2\pi )^{2}}\int du\,\,dt\sum\limits_{i,k}f_{i}f_{k}\frac{%
z_{i}^{2n}z_{k}^{2n}(1+\frac{z_{i}+z_{k}}{2})-\mu ^{2}}{t+1+uz_{i}+(1-u)z_{k}%
}.
\end{equation}
Numerical result (Table 1) depends very weakly on $n$ and is
compatible with the phenomenological value \cite{Nar}:
$\left\langle {\alpha }/{\pi }\;GG\right\rangle
=(393_{-38}^{+29}\;{\rm MeV})^{4}$.

\begin{table}
\caption{Condensates for $M=350$ MeV}
\begin{center}
\renewcommand{\arraystretch}{1.4}
\setlength\tabcolsep{5pt}
\begin{tabular}{cccccc}
\noalign{\smallskip} $n$& $ \Lambda$ &$\left\langle
\frac{\alpha }{\pi }GG\right\rangle$&
 $\left\langle \bar{q}q\right\rangle_{PS}$ &
 $\left\langle \bar{q}q\right\rangle_{PT}$
 & $\left\langle ig\,%
\bar{q}\,\sigma \cdot G\,q\right\rangle_{AV}$ \\
\noalign{\smallskip}
\hline
\noalign{\smallskip}
 $1$ & 1156 MeV & $(399~{\rm MeV})^4$ & $-(318~{\rm MeV})^3$
 & $-(357~{\rm MeV})^3$ & $-(553~{\rm MeV})^5$ \\
 $5$ & 2819 MeV & $(389~{\rm MeV})^4$ & $-(271~{\rm MeV})^3$
 & $-(301~{\rm MeV})^3$ & $-(475~{\rm MeV})^5$ \\
\end{tabular}
\end{center}
\label{Tab1a}
\end{table}

Soft pion theorems provide link between dynamical objects like the light
cone wave functions \cite{ChZh,BrLep,FarJack,EfRad} and static properties of
the physical vacuum \cite{ChZhitZhit,Zhitkt4}.
It has been shown in Refs.\cite{ChZhitZhit,Zhitkt4}
that moments of $\tilde{\phi}_{\pi }(k_{T}^{2})$ are given in terms of the
mixed quark-gluon condensates
\begin{equation}
\left\langle k_{T}^{2}\right\rangle _{AV}=\frac{5}{36}\frac{\left\langle ig\,%
\bar{q}\,\sigma \cdot G\,q\right\rangle }{\left\langle \bar{q}q\right\rangle
},\;\left\langle k_{T}^{2}\right\rangle _{PS}=\frac{1}{4}\frac{\left\langle
ig\,\bar{q}\,\sigma \cdot G\,q\right\rangle }{\left\langle \bar{q}%
q\right\rangle }.  \label{kt2}
\end{equation}
Here $G_{\mu \nu }^{a}$ is a gluon field strength and
$
\sigma \cdot G=\sigma _{\mu \nu }G^{\mu \nu }$,
$G_{\mu \nu }=\lambda^{a}/2 \; G_{\mu \nu }^{a}$.

Unfortunately the ratio $\left\langle k_{T}^{2}\right\rangle
_{AV}/\left\langle k_{T}^{2}\right\rangle _{PS}\sim 5/9$ which follow from (%
\ref{kt2}) is not reproduced within our approach\footnote{It is of the
order of 0.2 instead of 0.54} due to the slow convergence
of the $dk_{T}^{2}$ integration in the case of $\left\langle
k_{T}^{2}\right\rangle _{PS}$. In order to estimate the value of the mixed
condensate of dimension 5, $\left\langle ig\,\bar{q}\,\sigma \cdot
G\,q\right\rangle $, we choose therefore the first equation of Eqs.(%
\ref{kt2}).
Interestingly, for the parameters which are closest to
the original instanton model, $M=350$ MeV and $n=2$, we get $-(493$ MeV$)^{5}
$ in perfect agreement with the direct calculation of $\left\langle ig\,\bar{%
q}\,\sigma \cdot G\,q\right\rangle $ in the instanton model \cite
{Polyakov:1996kh} which gives $-(490$ MeV$)^{5}$.

\section{Summary and outlook}

The nonlocal NJL model with the momentum dependent constituent
quark mass has been applied to calculate pion light cone wave
functions \cite{Praszalowicz:2001wy}. It gives a satisfactory
description of the leading twist $AV$ 
wave function,
whereas for the twist 3 wave functions we find a somewhat larger
sensitivity to the model parameters.

Present prescription can be easily extended to describe kaon wave
functions with an explicit symmetry breaking due to the non zero
current strange quark mass. Also two meson generalized parton
distributions both for pions and kaons can be easily calculated.
By crossing symmetry one can also apply our method to calculate
the skewed distributions and structure functions \cite{ERA,Toki}.

On the theoretical side one has to investigate more closely the PCAC relation
within the present approach. It is known that the properly defined
currents should include additional terms with respect to those considered here
\cite{BroRipGol,nonl}. Although these new terms are not unique and suppressed
by the instanton packing fraction, their influence on our results
should be investigated.

\vspace{0.3cm}

M.P. thanks the organizers for the warm hospitality at this very
stimulating meeting.
This work was partially supported by the Polish KBN Grant PB~2~P03B~{\-}%
019~17. M.P. is grateful to W.Broniowski, K.Goeke, H.-Ch. Kim, D. M\"uller,
M.V.Polyakov and N.G.Stefanis for discussions and interesting suggestions.


\begin{thebibliography}{99}
\bibitem{ChZh}  V.L. Chernyak and A.R. Zhitnitsky, Sov. J. of Exp. and Theor.
Phys. Lett. \textbf{26} (1977) 359.

\bibitem{BrLep}  S. Brodsky and G.P. Lepage, Phys. Lett. \textbf{B87} (1979)
594; Phys. Rev \textbf{D22} (1980) 2157.

\bibitem{FarJack}  G. Farrar and D. Jackson, Phys. Rev. Lett. \textbf{43}
(1979) 246.

\bibitem{EfRad}  A.V. Efremov and A.V. Radyushkin, Theor. Mat. Phys. \textbf{%
42} (1980) 97; Phys. Lett. \textbf{B94} (1980) 245.

\bibitem{ChZhitZhit}  V.L. Chernyak, A.R. Zhitnitsky and I.R. Zhitnitsky
Sov. J. of Nucl. Phys. \textbf{38} (1983) 645 [Yad. Fiz. \textbf{38} (1983)
1074]. 

\bibitem{YakSch}  A. Schmedding and O. Yakovlev, Phys. Rev \textbf{D62}
(2000) 116002, [arXiv:hep-ph/9905392].

\bibitem {Cleo}J. Gronberg (CLEO Collaboration), Phys. Rev \textbf{D57} (1998) 33,
[arXiv:hep-ex/9707031].

\bibitem{ChZhPRep}  V.L. Chernyak and A.R. Zhitnitsky, Phys. Rep. \textbf{112%
} (1984) 173.

\bibitem{DPrev}  D.I. Diakonov and V.Yu. Petrov, {hep-ph/0009006} and
references therein.

\bibitem{DP}  D.I. Diakonov and V.Yu. Petrov, Nucl. Phys. \textbf{B245}
(1984) 259; \textbf{B272} (1986) 457.

\bibitem{Bochum}  V.Yu. Petrov and P.V. Pobylitsa, {\ hep-ph/9712203},
  V.Yu. Petrov, M.V. Polyakov, R. Ruskov, C. Weiss and K.
Goeke, Phys. Rev. \textbf{D59} (1999) 114018, [arXiv:hep-ph/9807229].

\bibitem{Praszalowicz:2001wy}  M.~Prasza\l{}owicz and
A.~Rostworowski,
Phys.\ Rev.\ D \textbf{64} (2001) 074003, [arXiv:hep-ph/0105188]
and hep-ph/0111196.


\bibitem{BrauFil2}  V.M. Braun and I.E. Filyanov, Z. Physik. {\bf C48}
(1990) 239.

\bibitem{Ball:1999je}  P.~Ball,
JHEP {\bf 9901} (1999) 010, [arXiv:hep-ph/9812375].


\bibitem{NiSKim}  N. G. Stefanis, W. Schroers and H.-Ch. Kim, Eur. Phys. J.
\textbf{C18} (2000) 137, [arXiv:hep-ph/0005218].

\bibitem{BaMiSte}  A.P. Bakulev, S.V. Mikhailov and N.G. Stefanis,
Phys.Lett. \textbf{B508} (2001), [arXiv:hep-ph/0103119]; hep-ph/0104290.

\bibitem{Nar}  S. Narison, Phys. Lett. \textbf{B361} (1995) 121,
[arXiv:hep-ph/9504334]; Phys. Lett. \textbf{B387} (1996) 162,
[arXiv:hep-ph/9512348].

\bibitem{Zhitkt4}  A.R.~Zhitnitsky, Phys. Lett. \textbf{B329} (1994) 493,
[arXiv: hep-ph/9401278];
A.R.~Zhitnitsky, Talk given at 10th Summer
School and Symposium on Nuclear Physics: QCD, Light cone Physics and Hadron
Phenomenology (NuSS 97), Seoul, Korea, 23-28 June 1997,
{hep-ph/9801228} and references therein.

\bibitem{Polyakov:1996kh}  M.~V.~Polyakov and C.~Weiss,
Phys.\ Lett.\ B \textbf{387} (1996) 841, [arXiv:hep-ph/9607244].

\bibitem{BroRipGol}  B. Golli, W. Broniowski and G. Ripka, Phys. Lett. {\
\textbf{B 437}} (1998) 24, [arXiv:hep-ph/9807261]; hep-ph/0107139.

\bibitem{nonl} R. S. Plant, M. C. Birse, Nucl. Phys. \textbf{A628} (1998) 607,
   R.D. Bowler, M. C. Birse, Nucl. Phys. \textbf{A582} (1995) 655,
   W. Broniowski, talk presented at the Miniworkshop on Hadrons as Solitons,
   Bled, Slovenia, 6-17 Jul 1999, hep-ph/9909438.

\bibitem{ERA}
R.M. Davidson, E. Ruiz Arriola, hep-ph/0110291;
Phys. Lett. \textbf{B348} (1995) 163;
H. Weigel, E. Ruiz Arriola, L.P. Gamberg, Nucl. Phys.\textbf{B560} (1999) 383,
[arXiv: hep-ph/9905329].

\bibitem{Toki}
T. Shigetani, K. Suzuki, H. Toki, Phys. Lett. \textbf{B308} (1993) 383 [arXiv:hep-ph/9402286];
Nucl. Phys. \textbf{A579} (1994) 413 [arXiv:hep-ph/9402277].

\end{thebibliography}
\end{document}